
---------------------------------------------------

\input harvmac

\def\bo{{\rm b_0}}\def\exp{{\rm exp}}

\def\DR{\overline{DR}}\def\MS{\overline{MS}}
\def\pf{{\rm Pf ~}}

\def\pf{{\rm Pf ~}}
\font\zfont = cmss10 
\def\bigone{\hbox{1\kern -.23em {\rm l}}}
\def\ZZ{\hbox{\zfont Z\kern-.4emZ}}

\noblackbox
\def\pf{{\rm Pf ~}}
\def\CO{{\cal O}}
\def\np#1#2#3{Nucl. Phys. B{#1} (#2) #3}
\def\pl#1#2#3{Phys. Lett. {#1}B (#2) #3}

\def\physrev#1#2#3{Phys. Rev. {D#1} (#2) #3}

\def\prep#1#2#3{Phys. Rep. {#1} (#2) #3}

\def\ev#1{\langle#1\rangle}

\Title{ \vbox{\baselineskip12pt\hbox{hep-th/9503115}
\hbox{RU-95-14 }\hbox{SLAC-PUB-95-6768}}}
{\vbox{\centerline{Instanton Calculations Versus }\smallskip
\centerline{Exact Results in 4 Dimensional SUSY Gauge Theories }}}
\bigskip\centerline{D. Finnell$^{1\dagger}$ and P. Pouliot$^2$}
\vglue .5cm\centerline{$^1$Stanford Linear Accelerator Center}
\centerline{Stanford University, Stanford, California 94309, USA}
\vglue .3cm\centerline{$^2$Department of Physics and Astronomy}
\centerline{Rutgers University, Piscataway, NJ 08855-0849, USA}
\bigskip\bigskip\noindent
We relate the non-perturbative exact results
in supersymmetry to perturbation theory using several different methods:
instanton calculations at weak or strong coupling,
a method using gaugino condensation and
another method relating strong and weak coupling.
This allows many precise
numerical checks of the consistency of these methods,
especially the amplitude of instanton effects,
and of the network of exact solutions in supersymmetry.
However, there remain
difficulties with the instanton computations at strong coupling.
\bigskip\bigskip\noindent
$^\dagger$Supported in part by the Department of Energy under
contract \#DE-AC03-76SF00515.\medskip\noindent\Date{3/95}
\newsec{Introduction}Instantons have played
an important conceptual role in
the study of non-perturbative effects in four dimensional
gauge theories.  However, their role in the quantitative
understanding of these theories has remained obscure.
In theories of physical interest like QCD, instanton calculations
are plagued by infrared divergences.
In the case of supersymmetric gauge theories,
non-renormalization theorems and
improved infrared properties allow one to make definite
statements about the effects of instantons.
Two main approaches have been taken to study
instanton effects in supersymmetric theories. Affleck, Dine and
Seiberg studied theories along the
flat directions of the scalar potential,
where the gauge group is spontaneously
broken and the theory is weakly coupled,
and computed the effective interactions of the light degrees of
freedom \ref\ads{I. Affleck, M. Dine and N. Seiberg,
\np{241}{1984}{493}; \np{256}{1985}{557}.}.
They demonstrated that instantons generate a calculable
superpotential in $SU(N_c)$ theories with $N_c-1$ flavors.
This weak coupling approach was adopted by the authors of
\ref\nsvzlargev{V.A. Novikov,
M.A. Shifman, A.I. Vainshtein and V.I. Zakharov, \np{260}{1985}{157}.}
who extended it to the study of Green functions and the
calculation of chiral condensates. Following \ads,
analytic continuation in holomorphic parameters was used to extend
the results obtained at weak coupling
to the strong coupling regime \ref\shiffvain{
M.A. Shifman, and A.I. Vainshtein, \np{296}{1988}{445}.}.
In the other approach, begun in
\ref\nsvz{V.A. Novikov, M.A. Shifman, A.I. Vainshtein and V.I.
Zakharov, \np{223}{1983}{445}; \np{229}{1983}{381, 394 and 407}.}
and continued in \nref\cernnpb{D. Amati, G.C. Rossi and G. Veneziano,
\np{249}{1985}{1}; D. Amati, Y. Meurice, G.C. Rossi and
G. Veneziano, \np{263}{1986}{591}.}\nref\cernrev{D. Amati,
K. Konishi, Y. Meurice, G.C. Rossi and G.
Veneziano, \prep{162}{1988}{169}.}\refs{\cernnpb,\cernrev},
the instanton calculations are performed
at the origin of field space in the strong
coupling regime. One computes chiral Green functions which
vanish in perturbation theory,
are independent of position by supersymmetry and are argued to
be saturated by instanton contributions;
cluster decomposition is then used
to extract the values of condensates.
Discrepancies arise in the quantitative comparison of the two
approaches \nref\germanfirst{
J. Fuchs and M.G. Schmidt, Z. Phys. C30 (1986) 161. }
\refs{\germanfirst,\shiffvain,\nsvzlargev}; there is a long-standing
controversy (section 9 of \cernrev)
over why this is so and which approach is correct.

\nref\nonren{N. Seiberg, \pl{318}{1993}{469}.}
\nref\nativacua{N. Seiberg, \physrev{49}{1994}{6857}.}
\nref\ils{K. Intriligator, R.G. Leigh and N. Seiberg,
\physrev{50}{1994}{1092}; K. Intriligator, \pl{336}{1994}{409}.}
\nref\ntwoi{N. Seiberg and E. Witten,
\np{426}{1994}{19}; erratum, \np{430}{1994}{485}.}
\nref\kn{K. Intriligator and N. Seiberg, \np{431}{1994}{551}.}
\nref\dual{N. Seiberg, \np{435}{1995}{129}.}
The new ingredient to this story is the progress over the last two years
in obtaining exact results about the low-energy behavior of
supersymmetric gauge theories \refs{\nonren - \dual}.
Following \ads, this recent progress allowed a greater number
of theories to be exactly related to one another.
The convenient way that is used to relate two different theories
is by expressing some non-perturbative
result of one theory in terms of the dynamically generated
scale $\Lambda$ of the other. Considering the results of\nref\ntwoii{N.
Seiberg and E. Witten, \np{431}{1994}{484}.}\nref\knati{K. Intriligator
and N. Seiberg, to appear.} \refs{\ads,\nonren - \knati}
as a whole, one sees an impressive web of interdependent
results. Sometimes it happens that in the solution of one
theory, a non-perturbative effect can directly be related
to perturbation theory. After this is done once, each subsequent
connection to perturbation theory provides an independent
consistency check of the network of exact solutions.

The next step in this framework, which is one purpose of this paper, is
to actually perform such checks with perturbation theory.
In section 2, we choose a specific perturbative regularization
scheme ($\DR$), define the scale of each theory in this scheme and relate
the scales of different theories in perturbation theory at one-loop.
To establish a first connection with perturbation theory,
we use a method involving an interplay between gluino condensation
and the strength of instanton effects in the $SO(4)$ theory with
one vector solved in \ils.
This connection precisely determines the amplitude
of non-perturbative effects for all the theories of the network.
By relating strong to weak coupling, the solution of N=2 $SU(2)$ SYM
of Seiberg and Witten provides another connection to perturbation theory,
fully consistent with the first one. Instanton calculations provide more
connections with perturbation theory. In the remaining sections,
we compare the results of explicit instanton calculations to
the predictions for their amplitude derived throughout section 2.
In section 3, we review the instanton calculations of
effective lagrangians, correcting some minor errors appearing
in the literature, and compare these results to the predictions.
In section 4, we compare the predictions to the Green functions
calculations in the weak coupling regime.  In both cases, we
find complete quantitative agreement. Finally, in section
5 we briefly point out some discrepancies when Green functions are
computed with instanton methods at strong coupling.
As a byproduct of this analysis, we obtain in section 2.4 the precise
normalization of the nonperturbative superpotentials for $SU(N)$
and gaugino condensates for $SU(N)$,
$SO(N)$ and $Sp(N)$ gauge groups.  This result should be
useful in practical applications of gaugino condensation,
e.g. in string phenomenology.
\newsec{Relating the scale $\Lambda$ to perturbation theory in $\DR$ }
\subsec{The $\DR$ regularization scheme}
In asymptotically free gauge theories, a scale $\Lambda$ is
dynamically generated by dimensional transmutation of the
gauge coupling and non-perturbative results are naturally
expressed in terms of this scale.
It is well-known that this scale can be defined in perturbation
theory, that it depends on the regularization
scheme chosen and that this dependence is exactly given
by a one-loop perturbative computation
\ref\collins{J. Collins, Renormalization,
Cambridge University Press, 1984.}.
We will find it convenient to use
the scheme of dimensional regularization through dimensional
reduction, with modified minimal subtraction, $\DR$.
This scheme was introduced by Siegel
\ref\siegel{W. Siegel, \pl{84}{1979}{193}.}
and it is standard in supersymmetry.
The coordinates and momenta are analytically continued
to $4-\epsilon$ dimensions, but it differs from the more common
$\MS$ scheme in that the number of components of spinor and
tensor fields is held fixed. Like the $\MS$
scheme, $\DR$ is a valid regularization scheme to
all orders in perturbation theory
\ref\jones{I. Jack, D.R.T. Jones, K.L. Roberts,
Z. Phys. C63 (1994) 151.}, but $\DR$ has the advantage of keeping
supersymmetry manifest. Of course, one could just as well work
in any other scheme, and the formulas
for translating to other conventional schemes such as $\MS$ and
Pauli-Villars are given below.

For a generic supersymmetric gauge theory, we define
the dynamical scale as
\eqn\scaledef{\Lambda^{\bo}_{\DR} \equiv \mu^{\bo}
e^{-8\pi^2/g^2_{\DR} (\mu)}}
with the bare coupling defined at a scale $\mu$.  Here $\bo$ is the
one-loop coefficient of the $\beta$ function, given by
$\bo={3\over 2}I_2(adjoint)-{1\over 2}\sum_R I_2(R)$,
where $I_2$ is the second Casimir index
and the sum is over the chiral multiplets in a representation $R$ of
the group (e.g. $\bo=3N_c-N_f-N_aN_c$ for
$SU(N)$ with $N_f$ flavors in the
fundamental and $N_a$ chiral multiplets in the adjoint representation).
This expression is invariant under
the renormalization group of the Wilsonian effective
action \nref\shifv{M.A. Shifman and A.I. Vainshtein,
\np{277}{1986}{456}; \np{359}{1991}{571}.}\nref\dines{M. Dine
and Y. Shirman, \physrev{50}{1994}{5389}.}\refs{\shifv,\dines}
to all orders in perturbation theory. It is not invariant under
the more usual renormalization group of the 1PI effective action,
however this difference appears at two and higher loops.
As stressed above, only a one-loop computation is required
for our purposes, and we will ignore this difference. In other
words, we will not write down explicitly the powers
of $g$ throughout this paper, as they are a higher loop effect.

\nref\hooft{G. 't Hooft, \physrev{14}{1976}{3432}.}
\nref\hasen{A.~ Hasenfratz and
P.~ Hasenfratz, \np{193}{1981}{210}.}
One more  advantage of $\DR$ is that it realizes the simple
threshold relations assumed in \ils.
The matching is most easily accomplished using background field
methods \refs{\hooft,\hasen}.
It involves computing the coefficient of the background field operator
$F^2_{\mu\nu}$ at one-loop and was done by Weinberg
\ref\wein{S. Weinberg, \pl{ 91}{1980} {51}.
L. Hall, \np{178} {1981} {75}.}.
When integrating out a set of superfields with the same canonical
mass $M$ (e.g. vector bosons and their superpartners
under the Higgs mechanism or heavy quarks and squarks
which decouple from the low-energy physics),
the formula of Weinberg expressed in the $\DR$
scheme yields the simple relation between the scales of the
low and the high energy theories:
\eqn\dec{ \Big({\Lambda_L \over M}\Big)^{\bo_L} =
\Big({\Lambda_H \over M}\Big)^{\bo_H}.}
That is, we find that the thresholds are
unity. In the case where the high-energy gauge group is broken,
the smaller group is assumed to be trivially embedded in the following
sense: we have chosen its generators to be a subset of
the generators of the larger group.
In other regularization schemes, such as $\MS$,
the relation \dec\ would be corrected
by finite numerical factors, which one should keep track of.
Relating the scale parameters in
different schemes is similarly done using the background field method.
One obtains a relation between the scales
of two different schemes A and B:
\eqn\translate{\Big({\Lambda^{(A)}\over \Lambda^{(B)}}\Big)^{\bo}=
 \exp\Big({I_2(adjoint)\over 2} (a_1^{(A)}-
a_1^{(B)}) - \sum_R {I_2(R)\over 2}(a_2^{(A)}-a_2^{(B)})\Big).}
In the $\DR$, Pauli-Villars and $\zeta$ function  schemes,
both $a_1$ and $a_2$ vanish, while
for $\MS$, $a_1=1/6$ and $a_2=0$.  We thus have the
well-known relations between schemes
\nref\vaughn{S.P. Martin and M.T. Vaughn,
\pl{318}{1993}{331}.}\nref\rebhan{A. Rebhan,
\physrev{39}{1989}{3101}.}\refs{\hasen,\vaughn,\rebhan}:
\eqn\schemes{\Lambda_{\DR}^{\bo}=
\Lambda_{PV}^{\bo}=\Lambda_{\zeta}^{\bo}=
e^{-I_2(adjoint)/12}\Lambda_{\MS}^{\bo}.}
\subsec{Matching perturbation theory to the exact solutions}
In this section, we will show that the exact solution of
the $SO(4)=SU(2)_1\times SU(2)_2$ theory with
matter $Q$ in the vector $(2,2)$ representation
provides a direct way to relate the non-perturbative results
to perturbation theory. The effective superpotential for this model was
exactly determined by Intriligator, Leigh and Seiberg in \ils\
up to an arbitrary constant $c$:
\eqn\superpotentialils{
W= S_1 \Big[\log \Big({c\Lambda_1^5\over S_1^2}\Big)+1\Big] +
S_2 \Big[\log \Big({c\Lambda_2^5\over S_2^2}\Big)+1\Big] +
(S_1+S_2) \log\Big({S_1+S_2 \over X}\Big) + mX }
where $X=\det Q=Q_1^\alpha Q_{2\alpha}$
and $S_i = {W_{i\alpha}^{a2}\over 32\pi^2}$ are the glueball
superfields. The following argument shows
that, for consistency of the model, we
must have $c=1$ in the $\DR$ scheme.
For $m\neq 0$, integrating out $X$ leads to two pure
$SU(2)$ gauge theories with
$W= 2 S_1 + 2 S_2$.  Solving for $S_i$ gives
$S_i = \pm \sqrt c \tilde\Lambda_i^3$
where $\tilde\Lambda_i^3$ is
the scale of the low-energy $SU(2)_i$ group and
$\tilde\Lambda^6_i= m\Lambda^5_i$ from
equation \dec. This shows that with
this set of conventions and normalizations, a contribution of
$2S$ to the superpotential must appear
with $S= \pm \sqrt c \Lambda^3_{0,0}$
whenever gaugino condensation occurs.
If we now take $X$ to be massless  and take $\ev X\neq 0$,
the gauge group is broken to a diagonal subgroup $SU(2)_D$.
This is a pure $SU(2)$ SYM theory up to irrelevant interactions
with the field $X$. Therefore we expect that
$S_D=\pm \sqrt c \Lambda_D^3$, where the matching
$\Lambda_D^3 = \Lambda_1^{5/2}
\Lambda_2^{5/2}/\ev{X}$ was performed using equation \dec\ and
recalling that $1/g_1^2 + 1/g_2^2 = 1/g_D^2$.
Integrating out $S_1$ and $S_2$ from
equation \superpotentialils\ gives $W_{eff}=c (\Lambda_1^{5/2}\pm
\Lambda_2^{5/2})^2/X$. The cross-term
$\pm 2 c \Lambda_1^{5/2}\Lambda_2^{5/2}/X$ must arise from
gluino condensation in $SU(2)_D$. For this to
be the case, we must have
$\sqrt c =c$. Therefore $c=1$. To summarize, taking the coupling
$g_2=0$, we deduce that the gluino condensate of pure $SU(2)$ SYM is
\eqn\puregluino{\ev S = \pm \Lambda^3}
and that the superpotential of the $SU(2)$
theory with one flavor is \eqn\oneflavor{{\Lambda^5\over X}.}
Thus, the exact solutions predict that the coefficient of the
well-known instanton-induced superpotential
of Affleck, Dine and Seiberg is 1 when
expressed in terms of $\Lambda_{\DR}$.
\subsec{$SU(2)$ model with 2 doublets and a triplet }
The solution of this theory by Intriligator and Seiberg
\kn\ relates the non-perturbative results of the $SU(2)$
theories without matter,
with one fundamental flavor or with one adjoint.
Our purpose is to match this solution to perturbation
theory via the connection established in section 2.2.
We give some details to illustrate precisely what we mean.
This is straightforward and, for other models,
we will simply quote in section 2.4 the results of such matchings.

We begin by taking in its entirety the solution of the model in \kn.
We will only summarize the features that we need.
The basic gauge singlets are $X=Q^\alpha_1Q_{2\alpha}$, $U=
{1\over 2}\phi^a\phi^a$  and $\vec Z={\sqrt 2\over 2} Q^\alpha_f
\phi^b \sigma^b_{\alpha\beta} Q^\beta_g
\vec\sigma^{fg}$ ($\alpha,\beta,f,g=1,2; a,b=1,2,3$).
Their expectation values label the inequivalent vacua.
They are constrained classically by $UX^2=\vec Z^2$,
but they are independent quantum mechanically.
It is enough for our purposes
to consider only the subspace $\vec Z=0$. We will use
the notation $\Lambda_{N_f,N_a}$ for the scales of the theories
obtained by reducing the number of fundamentals (of mass $m$) or adjoints
(of mass $M$) by giving masses to them and integrating them out.
The matching of the scales is simply $mM^2\Lambda^3_{1,1} =
M^2\Lambda^4_{0,1}
= m\Lambda^5_{1,0}=\Lambda^6_{0,0}$ using equation \dec.
This model has three phases:
\item{1. } For generic VEVs of $X, U$, the model is in a
Higgs-confining phase described
by the superpotential (whose form is determined by the symmetries)
\eqn\superpotential{W={-XU^2\over 4\Lambda^3_{1,1} }.}
It describes the theory everywhere away from the
Coulomb phase. The normalization of this
superpotential is chosen in order to match the
results of the previous section, as will be shown below.
\item{2. } On the line labeled by $U\neq 0$,
$X=0$, the theory is in a free Coulomb phase.
The quantum moduli space is singular on this line,
because the elementary fields $Q_f$ are charged under an unbroken
$U(1)$. \item{3. } At $X=U=0$, the  theory is
in an interacting non-Abelian Coulomb phase.

Now we consider perturbing this theory by mass terms.
Adding $MU$ to the superpotential \superpotential\
and integrating out $U$, we find an
effective $SU(2)$ theory
with one fundamental flavor, with precisely
the superpotential \oneflavor. Thus we have the chosen the
correct normalization in \superpotential.
Adding a mass $mX$ to the superpotential
\superpotential\ and integrating out $X$,
we flow to the N=2 $SU(2)$ SYM model
(for more details see section 2.5), at the location of its
singularities on its quantum moduli space:
\eqn\location{\ev U = \pm 2 \Lambda_{0,1}^2.}

Integrating in $S$, starting from \superpotential, leads to
$W=S[\log\Big( {4\Lambda^3_{1,1} S\over XU^2}\Big) - 1]$.
We give a mass to $X$ and integrate it out, to get
$W=S \log\Big( {4m\Lambda^3_{1,1} \over U^2} \Big)$.
Giving a mass to $U$ and integrating it out yields
$W= S[\log\Big( {mM^2\Lambda^3_{1,1} \over S^2} \Big) +2]$:
we have flown to a pure $SU(2)$ SYM theory and
recovered that gaugino condensation occurs at the location \puregluino:
\eqn\confinings{\ev S = \pm \Lambda^3_{0,0}.}
We view this as a consistency check on the matching of the $SU(2)\times
SU(2)$ model with a $(2,2)$ to the $SU(2)$ model with one flavor
and one adjoint. Along the way, we also got that in the confining phase:
\eqn\massstuff{m\ev X=\ev S \qquad {\rm and} \qquad M\ev U=2\ev S.}
These relations between condensates are seen to follow from
the exact solution of Intriligator and Seiberg. Alternatively,
they are known to arise from the Konishi anomaly
\ref\ggrs{S.J. Gates, Jr., M. Grisaru, M. Rocek and W.
Siegel, Superspace or One Thousand and One
Lessons in Supersymmetry", Benjamin/Cummings, Reading 1983.  K. Konishi,
\pl{135}{1984}{439}.}.
\subsec{Other models} Using the result for $SU(2)$
with one flavor obtained in equation \oneflavor, we fix the normalization
of the superpotentials in $SU(N_c)$ theories with $N_f< N_c$
flavors by induction. We start with $SU(N_c)$ with $N_c-1$ flavors,
by giving a large VEV to one flavor, and flow to $SU(N_c-1)$
with $N_c-2$ flavors.  Repeating this procedure, we are eventually
lead back to $SU(2)$ with one flavor.  Having fixed the
superpotential for $N_f=N_c-1$, we then add mass terms to
flow to theories with fewer flavors.
Using this method \refs{\ads,\nativacua}, we obtain
\eqn\dynsuper{W_{eff} = (N_c-N_f) {\Lambda^{3N_c-N_f \over
N_c-N_f}\over (\det Q \tilde Q)^{1\over N_c-N_f} }, }
for $N_f<N_c$, exactly as in \nativacua,
but now $\Lambda$ has been precisely identified with
$\Lambda_{\DR}$. For the case of $N_f=N_c$, there is no superpotential,
but \nativacua\ the classical constraints
$\ev{\det M-B\tilde B}=0$ ($\ev {\pf V}=0$ for $SU(2)$)
are modified quantum mechanically to
\eqn\constraints{\ev{\pf V} = \Lambda^4
\qquad {\rm and} \qquad \ev{\det M-B\tilde B} =
\Lambda^{2N_c}. }  Again, we have fixed the
normalization by adding a mass and flowing to
the $N_f=N_c-1$ theory, and used the notation
$V_{ij}=Q_iQ_j$, $\pf V = {1\over 8}\epsilon_{ijkl}
V^{ij}V^{kl}$ for $SU(2)$ and $M_{i\tilde{\jmath}}=Q_i\tilde
Q_{\tilde{\jmath}}$ and
$B=Q_1Q_2\cdots Q_{N_c}$ for $SU(N_c)$ (with color indices suppressed).

We obtain the gaugino condensates of other groups
in a similar way \nref\russiangluino{
A.Yu. Morozov, M.A. Olshanetsky and
M.A. Shifman, \np{304}{1988}{291}.}\refs{\ads,\shiffvain,\russiangluino}:
and obtain (where we have taken both scales equal for $SO(4)$):
\eqn\condensaten{\ev{{\lambda^a\lambda^a\over 32\pi^2}}_{SU(N)}=
\Lambda^3, }\eqn\condensates{\ev{{\lambda^a\lambda^a\over
16\pi^2}}_{SO(N), N\ge 4}= 2^{4/(N-2)}\Lambda^3 \qquad {\rm and} \qquad
\ev{{\lambda^a\lambda^a\over 64\pi^2}}_{Sp(2N)}=2^{-2/(N+1)}
\Lambda^3.} To derive this result for $Sp(2N)$,
one has to take into account in using equation
\dec\ that when $Sp(2N)\to Sp(2N-2)$ by the VEV $v$ of a flavor of
fundamentals, the vector bosons which are singlets under $Sp(2N-2)$
have mass $v$, while those who transform as the $2N-2$
representation have mass $v/\sqrt 2$.
Although this will not be needed here, other models are easily
matched to the model of section 2.2. Namely, all the results of \ils\ for
$SO(5)\times SU(2)$ and $SU(2)\times SU(2)$ models with various matter
contents have already the correct normalization for the $\DR$ scheme.
For the $SU(2)$ theory with 2 triplets solved in \kn, a
rescaling $\Lambda\to 4\Lambda_{\DR}$ is required and, similarly, for the
$N=2$ $SU(2)$ theories with $i$ matter hypermultiplets solved in
\ntwoii, the rescaling should be $\Lambda^{4-i}_i \to
4 (\Lambda^{4-i}_i)_{\DR}$.
\subsec{A second method to relate exact solutions to perturbation theory}
In this section we relate the solution of N=2 $SU(2)$ SYM
obtained by Seiberg and Witten \ntwoi\
to perturbation theory.  This should be viewed as an alternate
method to that of section 2.2. Before jumping into technical details,
let us first outline the procedure.
In terms of N=1 components, the theory contains a vector multiplet
$W_\alpha^a= (\lambda_{\alpha}^a,F_{\mu\nu}^a)$
and a chiral multiplet $\Phi^a=(\phi^a,\psi^a_{\alpha})$
in the adjoint representation. N=2 SUSY is unbroken and
forbids any superpotential.  Thus, the adjoint $\phi^a$ may obtain a
complex VEV $(0,0,a)$, breaking the $SU(2)$ down to $U(1)$.
The low energy degrees of freedom are then the $U(1)$ photon
multiplet and a neutral chiral superfield
$a = a +\sqrt 2 \theta \psi +\theta^2 F$.
(We will use the same notation $a$, and $U={1\over 2}\phi^a\phi^a$
needed later, for the chiral superfields, their lowest component,
and the VEV of their lowest component.)
The low-energy effective lagrangian for these fields contains
a kinetic term for $a$ and an effective gauge coupling
\eqn\taua{{1\over 16\pi} Im \int d^2\theta \tau(a) W_\alpha^2.}
If the VEV $a$ is large, the theory is semiclassical, and the general
form of $\tau$ follows from the symmetries
\ref\nati{N. Seiberg, \pl{206}{1988}{75}.}
\eqn\taub{\tau(a)={2i\over\pi}\log{C_0 a^2\over\Lambda^2}+
\sum_{l=1}^\infty {C_l \Lambda^{4l}\over a^{4l}}.}
The logarithmic term is just the perturbative renormalization
of the coupling constant
and the power corrections are generated by instantons.
By relating strong coupling to weak coupling,
the solution of Seiberg and Witten gives
all the coefficients $C_l$ exactly.
The scale $\Lambda$ can be matched to $\DR$ by comparing this
logarithmic term to that calculated directly
in the $\DR$ scheme using
equation \dec.  This requires that we extract the constant
$C_0$ from the exact solution for $\tau$.
We will also extract the coefficient $C_1$ of the one-instanton
term for comparison with an explicit instanton calculation.

We now give a brief description of the exact solution and
the steps needed to extract this information.
The VEV $a$ is not a good global coordinate
on the quantum moduli space, and is
traded for $U$ defined above, which is.
The low energy effective lagrangian is expressed in terms
of two functions of $U$, $a=a(U)$ and $a_D(U)$
\eqn\lagrangian{L={1\over 8\pi} Im \int d^4\theta a_D \bar a +
{1\over 16\pi} Im \int d^2\theta \tau(U) W_\alpha^2 }
with $\tau ={\theta\over \pi}+ {8\pi i \over e^2} = a'_D/ a'$,
and $\theta(U)$ and $e(U)$ the $U(1)$ effective couplings
(note that we are using for the normalization of
$\tau$ the conventions of \ntwoii\
which differ from \ntwoi\ by a factor of 2). The exact solution
gives $a(U)$ and $a_D(U)$ as periods of the elliptic curve
\eqn\ellipticcurve{y^2=x^3-U x^2 +\Lambda^4 x,}
with the same conventions as in the previous sections.
This formula is the definition of $\Lambda$.
The curve is singular when $U=\pm 2\Lambda^2$ and
is branched at $x=\infty$, $x_0=0$ and
$x_\pm = {1\over 2}(U\pm \sqrt{ U^2-4\Lambda^4})$.
Seiberg and Witten found that \eqn\adual{ a'_D = {da_D\over dU}=
c_1 \int_{x_-}^{x_+} {dx\over y}\qquad a'={da\over dU}=
c_1 \int_{x_0}^{x_-} {dx\over y} } with $c_1$ a constant,
which leads to the correct monodromies
\eqn\monodromies{\pmatrix{1 & 0 \cr -1 & 1 \cr} {\rm \ at\ }
U\sim 2\Lambda^2 {\rm \ and} \qquad
\pmatrix{-1 & 4 \cr 0 & -1 \cr} {\rm at\ large\ U}. }
acting on the vector $(a'_D,a')$.
Equations \adual\ have the explicit solution
\eqn\elliptic{a'_D= i {c_2 \over \sqrt{x_+}} K(k') \qquad
a'= {c_2 \over \sqrt{x_+}} K(k) }
where $K$ is the elliptic function of the first kind,
$k=\sqrt{x_-/ x_+}$ and $k'=\sqrt{1-k^2}$
and $c_2$ is a constant. The limit $U\rightarrow \infty$
corresponds to $k\rightarrow 0$. We make use of the expansion
\eqn\ellipticexp{K(k) = \cases{  {\pi\over 2} (1+{k^2\over 4}+\CO(k^4)),
&$k\rightarrow 0$; \cr (1+{k'^2\over 4})\log ({4\over k'})
-{k'^2\over 4} +\CO(k'^4),&$k\rightarrow 1$.\cr} }
At large $U$, the theory is semiclassical and we find
\eqn\finaltau{\tau=a'_D/a'\cong {2i\over \pi}\log{4 U \over \Lambda^2}.}
For the matching between
$SU(2)$ and $U(1)$, with the mass of the heavy
bosons $W_\pm$ being $2\sqrt U$ classically, equation \dec\ yields
${-8\pi^2\over e^2}= {-8\pi^2\over g^2(2\sqrt U)} =
-4\log( {2\sqrt U\over \Lambda_{\DR}})$
where $g$ is the $SU(2)$ coupling.
But from equation \finaltau, $Re\ [i\pi\tau]={-8\pi^2\over e^2}=
-2 \log({4 U\over \Lambda^2})$.
Thus we identify $\Lambda$ with $\Lambda_{\overline{DR}}$ which is one
result of this section. Moreover, the location of the singularities
$U=\pm 2\Lambda^2$ in $\DR$, matches correctly with equation \location.

More work is needed to extract the one-instanton predictions
of the effective lagrangian \lagrangian. We will be interested
in the four-fermi interaction $\psi\psi \lambda\lambda$ predicted
by the photon kinetic term. To the one-instanton approximation,
\eqn\uinst{
U= {a^2\over 2}(1+2 {\Lambda^4\over a^4}+\CO({\Lambda^8\over a^8}) ),}
obtained by integrating $a'$ with respect to $U$.
One gets different formulas for $\tau$ depending
on whether one expresses it as a function of $a$ or of $U$.
For comparison with instanton calculations, it is appropriate to
express $\tau$ as a function of $a$;
\eqn\tauofu{\tau(a) = {2i\over \pi}\Big( \log{2 a^2 \over \Lambda^2}
-{ 3\Lambda^4 \over a^4}+\CO({\Lambda^8\over a^8})\Big).}
To get the four-fermion interaction, remember that $a$ above
is a chiral superfield, so that
$$a^{-4}(\theta) = a^{-4} (1+{\sqrt 2 \theta\psi \over a})^{-4}
= a^{-4} (1-10{\theta^2 \psi^2\over a^2}) +other\ terms.$$
Therefore, we get the prediction expressed in terms of $\tau(a)$:
\eqn\effectiveinteraction{S_{eff} \supset {1\over 16\pi}
Im\int \tau(a) W_\alpha^2 d^2\theta d^4x=
{15\Lambda^4\over 8\pi^2 a^6} \int d^4x
\psi^2\lambda^2 +h.c.+other\ interactions. }
\newsec{Instanton Calculations of $L_{eff}$} For $SU(N_c)$ with
$N_c-1$ flavors, the superpotential \dynsuper\ leads to a mass term
for a classically massless fermion.
Affleck, Dine and Seiberg \ads\ showed
that this dynamical mass is generated by instantons.
The numerical coefficient was
computed by Cordes in \ref\cordes{S. Cordes, \np {273} {1986}{ 629}.}.
The effective lagrangian \effectiveinteraction\ in the N=2 theory
contains a four-fermi interaction of massless fermions, also generated
by an instanton amplitude; Seiberg \nati\ showed
that the coefficient is non-zero. We will show that both calculations
agree quantitatively with the exact predictions.

Because of supersymmetry, the nontrivial eigenvalues which enter
the instanton determinant cancel between boson and fermion sectors.
Thus, such instanton calculations reduce to tree level perturbation
theory and  combinatorics of fermion zero modes.
Both these aspects are considerably simplified in the
superfield formalism of \refs{\nsvz,\nsvzlargev}.
Here we will work in components. We assume much familiarity with
standard instanton notation \hooft\
and we note that our normalization of the instanton amplitude
agrees entirely with the corrected calculational framework
summarized in \ref\morethooft{G. 't Hooft, \prep{142}{1986}{357}.}.
\subsec{ The computation of $W_{eff}$ in $SU(2)$ with one flavor}
We present a simplified version of the calculation of Cordes in \cordes,
specific to the case of $SU(2)$.
We are concerned with getting precisely the numerical coefficient
that determines the instanton amplitude. This is determined by one-loop
fluctuations about the instanton. This computation assumes a large
vacuum expectation value $\ev {Q^i_f}=v\delta^i_f$
along the flat direction
of the classical potential for
the two  matter doublets $Q_f=Q_f+\sqrt 2 \theta\psi_f +\theta^2 F_f$,
($i$ is the color index). In the instanton background, this becomes
$$Q^i_f  = { \sigma^{\mu i}_f x_\mu v \over \sqrt{ x^2+\rho^2} }.$$
This VEV provides an infrared cutoff, which suppresses
large instantons exponentially, since the classical action is now
$8\pi^2/g^2+4\pi^2\rho^2 v^2/g^2$.
Along this flat direction there is a classically massless fermion
$\chi_\alpha = {1\over \sqrt2}(\psi_1^1 + \psi_2^2)_\alpha$,
($\alpha$ is the spin index)
which obtains a mass from the superpotential \oneflavor.
The considerations of the previous section predict this mass.
Expanding the superpotential \eqn\superchi{ {\Lambda^5 \over Q_1Q_2} =
- {\Lambda^5\over v^4} ({3\over 4}(\psi_1^1+\psi_2^2)^2 +
{1\over 4}(\psi_1^1-\psi_2^2)^2 ) + \cdots =
 {3 \over 2} {\Lambda^5 \over v^4} \chi\chi+other\ terms. }
This gives $m_\chi = 3\Lambda^5/v^4$.

We now do the instanton calculation of $m_\chi$.
The bosonic zero modes and the
classical action give a standard factor (see, e.g.,
\refs{\hooft,\cernrev}):
\eqn\bosonic{2^{10}\pi^6  \int {d^4x_0 d\rho\over \rho^5}
e^{-{8\pi^2/ g^2(\mu)}} (\mu\rho)^8 e^{-4\pi^2\rho^2 v^2} }
The precise definition of $g^2(\mu)$ in the exponential depends
upon how one chooses to regulate the instanton determinants.
It is easiest to use $\zeta$-function
or alternatively Pauli-Villars \hooft\ regularization.
As explained in section 2.1, for either choice the resulting $\Lambda$
scale can be identified with $\Lambda_{\DR}$.

At zeroth order in $\rho v$,
there are 6 fermion zero modes.
They are normalized according to
$\Sigma_k \int \psi^*_k(x) \psi_k(x) d^4x=1$,
with $k$ representing color and Lorentz indices.
The zero modes form three pairs: \eqn\fermionic{
\lambda_\alpha^{SSa[\beta]}= {\sqrt 2 \rho^2\over \pi}
{\sigma_\alpha^{a\beta} \over (x^2+\rho^2)^2} \qquad
\lambda_\alpha^{SCa[\dot\beta]}= {\rho\over \pi} {x_\beta^{\dot\beta}
\sigma_\alpha^{a\beta} \over (x^2+\rho^2)^2} \qquad
\psi_{\alpha f}^{i[n]} = {\rho\over \pi}
{\delta_f^n \delta_\alpha^i \over (x^2+\rho^2)^{3/2} }. }
The indices in brackets label the two members 1,2 of a pair;
the explicit indices are $\alpha$ for spin,
$i, a$ for color and $f$ for the two doublets.

The Yukawa coupling $\lambda\psi Q^{\dag}$ perturbs this picture
qualitatively.  First
it lifts the superconformal and matter fermion zero modes
(for details, see \cordes), leaving the factor \eqn\grassmannian{
\mu^{-2}\int d\xi_1 d\xi_2 d\tau_1 d\tau_2\exp( {i\sqrt 2 }
\int d^4x \xi \lambda^a T^a Q^{\dag} \psi \tau )= {v^2\over 2\mu^2}.}
Second, it mixes the massless fermion
with the supersymmetric zero modes. The classical equations are
$\rlap/ D\lambda = 0$ and $\rlap/ D^{\dag} \psi^{\dag} =
\sqrt 2 Q^{\dag} T^a\lambda^a$
which have the solution (up to negligible higher order corrections
in $(\rho v)^2$)\eqn\solution{\lambda=\lambda_{SS} \qquad
\psi^{{\dag} i[\beta]}_{\dot\alpha f} =
{1\over 4\pi} (\rlap/ D^{\beta}_{\dot\alpha })_j^i
 Q^{\dag j}_f . } {}From \solution, we get
the classical wave-function of $\chi^{\dag [\beta]}_{\dot\alpha}$,
with an index $[\beta]$ telling which supersymmetric
zero mode it is attached to. We restore the instanton location $x_0$,
and take $|x-x_0|\to \infty$.  In this limit
\eqn\chidagger{\chi^{\dag [\beta]}_{\dot\alpha}
 = {1\over 4\pi\sqrt2}(( \rlap/ D_{\dot\alpha}^\beta)^1_j Q^{\dag j}_1 +
 (\rlap/ D_{\dot\alpha}^\beta)^2_j Q^{\dag j}_2)
\simeq {v \rho^2 \over 4\sqrt 2 \pi} \rlap/
 \partial_{\dot\alpha}^\beta {1\over (x-x_0)^2}
= {\pi\rho^2 v\over \sqrt 2} S_{F\dot\alpha}^\beta (x,x_0) }
where $S^\beta_{F\dot\alpha}$ is the standard free massless
Feynman propagator: $$4\pi^2 S_{F\dot\alpha}^\beta (x,x_0) \equiv
\rlap/ \partial_{\dot\alpha}^\beta {1\over (x-x_0)^2}.$$
Finally, collecting the above factors, we compute a Green function
for large $|x-y|$: \eqn\mass{\eqalign{
\ev{\chi^{\dag}_{\dot\alpha} (x) \chi^{\dag\dot\alpha}(y)}= &
2^{10}\pi^6\int d^4x_0d\rho \rho^3\Lambda^5 \Big({v^2\over2}\Big)
\Big({\pi\rho^2 v\over \sqrt2}\Big)^2e^{-4\pi^2 v^2\rho^2}
S_{F\alpha}^{\dot\beta}(x,x_0)
S_{F\dot\beta}^{\alpha}(x_0,y) \cr =&
3{\Lambda^5\over v^4}\int d^4x_0 S_{F\alpha}^{\dot\beta}(x,x_0)
S_{F\dot\beta}^{\alpha}(x_0,y) .}}
Comparing with the amplitude generated by a mass term \eqn\wmass{
\ev{\chi^{\dag}_{\dot\alpha} (x) \chi^{\dag\dot\alpha}(y)}
= m_\chi  \int d^4x_0 S_{F\alpha}^{\dot\beta}(x,x_0)
S_{F\dot\beta}^{\alpha}(x_0,y) .}
we find complete agreement with the prediction equation \superchi.

This calculation was carried out by Cordes for any number of colors.
It involves a nontrivial integration over the orientation of
the instanton.  Note that the result in \cordes\ disagrees with
the prediction \dynsuper\ by a factor of 4 as the result of some minor
errors\foot{In particular, the extraction of the mass term, equation
6.9, and multiplicative errors appearing in equations 5.19 and 6.7.}
appearing in that paper.  Correcting for these, the result
for arbitrary $SU(N)$ with $N-1$ flavors is
\eqn\wsun{W_{SU(N)}={\Lambda^{2N+1}\over {\rm det\ } Q \tilde{Q}}}
in precise agreement with the prediction \dynsuper.
\subsec{The four-fermi interaction, equation \effectiveinteraction}
The calculation of the four-fermi interaction in the N=2 $SU(2)$ SYM
theory is very similar to that of the previous section \nati\ and we
will be very brief. The gauge group is $SU(2)$ and there is
a chiral superfield in the adjoint
$\phi^a= \phi^a + \sqrt 2 \theta \psi^a+\theta^2 F^a$.
The scalar zero mode is: \eqn\scalarstructure{
\phi^a_{instanton}= {a x_\mu x_\nu \eta^a_{\mu\lambda}
\bar\eta^b_{\nu\lambda} V^b\over x^2+\rho^2} }
where $V^b= (0,0,1)$ is a chosen direction and $a$ is the VEV.
The group is broken to $U(1)$, leaving
a massless $N=2$ vector multiplet.  There are now two classically
massless fermions, which we label $\psi=\psi^3$ and $\lambda=\lambda^3$.
There are 8 zero modes; the 4 gaugino zero modes of
\fermionic, and identical supersymmetry and superconformal zero modes
associated with $\psi^a$. The Yukawa coupling $\sqrt 2
\epsilon^{abc}\psi^a\lambda^b \phi^{\dag c}$
raises the 4 superconformal zero modes, leaving a factor
of $a^2/2\mu^2$ and mixes the massless fermion
$\lambda^\dagger$ with $\psi_{SS}$ and mixes $\psi^\dagger$ with
$\lambda_{SS}$. Far from the instanton location, the
 classical wavefunctions of the massless fermions are
$\psi^{\dag\dot\alpha}_{[\alpha]},
 \lambda^{\dag\dot\alpha}_{[\alpha]}=
\pi\rho^2 a S_{F\alpha}^{\dot\alpha}$.
The four-fermion amplitude computed at large separation is
\eqn\green{\eqalign{\ev{\psi^{\dag\dot\alpha}(x_1)
\psi^{\dag}_{\dot\alpha}(x_2)\lambda^{\dag\dot\beta}(x_3)
\lambda^{\dag}_{\dot\beta}(x_4)}=&
2^{10}\pi^6\int d^4x_0 d\rho \rho^3 \Lambda^4 \Big({a^2\over 2}\Big)
(\pi\rho^2a)^4 e^{-4\pi^2\rho^2 a^2}\cr
&S^{\dot\alpha}_{F\alpha}(x_1,x_0)S^{\alpha}_{F\dot\alpha}(x_0,x_1)
S^{\dot\beta}_{F\beta}(x_3,x_0)S^{\beta}_{F\dot\beta}(x_0,x_4) \cr
={15 \Lambda^4\over 2\pi^2 a^6}\int d^4 x_0 &
S^{\dot\alpha}_{F\alpha}(x_1,x_0)S^{\alpha}_{F\dot\alpha}(x_0,x_1)
S^{\dot\beta}_{F\beta}(x_3,x_0)S^{\beta}_{F\dot\beta}(x_0,x_4)}}
which is exactly the amplitude predicted by the effective interaction
\effectiveinteraction.
\newsec{Green functions at weak coupling}
Another fruitful approach to studying SUSY theories has
been the instanton computation of chiral Green functions.
Supersymmetry guarantees that correlation functions of lowest
components of chiral superfields, which vanish in perturbation
theory, will be independent of position.
If the Green functions can be computed reliably,  then clustering
can be used to extract the values of various gauge invariant
condensates. In this section, we compare the calculations
done at weak coupling in the formalism of
\refs{\nsvzlargev,\germanfirst}
with the predictions.
We differ from the literature \nref\germansecond{
J. Fuchs, \np{272}{1986}{677}.}\refs{\cernrev,\germanfirst,\germansecond}
which uses zero modes normalized according to
$\int{\rm tr}\lambda\lambda=1$ for fermions in the adjoint
representation, whereas we take
$\int\sum_a \lambda^a\lambda^a=2\int{\rm tr}\lambda\lambda=1$ in
accord with \fermionic.
Results quoted in the literature must therefore be corrected by
factors of $1/2$ for every pair of zero modes in the adjoint
representation.

In the $SU(2)$ theory with one massless flavor,
assuming a large VEV for $X$ classically,
one computes:
\eqn\greensint{ \ev{S(x_1) X(x_2)} = \Lambda^5.}
To compare this with the exact solution, we use the effective
superpotential of equation \oneflavor, and add a mass term $mX$.
Integrating out $X$, we find $\ev X=(\Lambda^5/m)^{1\over2}$. Using the
relation $\ev S=m \ev X$ of equation \massstuff,
we get $\ev {SX}= \ev S \ev X=\Lambda^5$, which
agrees with the explicit calculation \greensint.

In the $SU(2)$ theory with 2 massless flavors, one computes
at weak coupling (with a large VEV for $V_{12}$ for example):
\eqn\twoflavors{ \ev{{1\over 8}\epsilon^{ijkl}
V_{ij}(x_1) V_{kl}(x_2)} = \Lambda^4 }
which agrees with the prediction \constraints.

Finally, in the N=2 SU(2) SYM theory, Green function methods
at weak coupling allow us to compute
the quantum correction to the classical
relation $\ev U\cong{1\over2}\ev{a^2}$
at large $a$. One computes,
\eqn\uu{\ev {U(x)}={1\over2}a^2+
{\Lambda^4\over a^2}+\CO({\Lambda^8\over a^6})}
in perfect agreement with the prediction \uinst, and \eqn\uuu{
\ev{U(x_1) U(x_2)}={1\over4}a^4+\Lambda^4+\CO({\Lambda^8 \over a^4}),}
in agreement with the clustering property.
\newsec{Green functions at strong coupling}
In another approach to instanton calculations,
summarized in \cernrev, Green functions of gauge invariant,
chiral operators are computed directly in the confining phase, without
vacuum expectation values of the elementary fields.
Instantons are expected to saturate the amplitude and
the integration over the instanton size is convergent,
due to the finite separation between the operators inserted.
Using the position independence guaranteed by supersymmetry,
and cluster decomposition, the values of
the condensates can be extracted.

In the case of pure $SU(2)$, one computes
in the formalism of \refs{\cernrev,\germanfirst}:
\eqn\puresutwo{\ev{S(x_1) S(x_2)} = {4\over 5}\Lambda^6.}
{}From this, one extracts $\ev S=\pm\sqrt{4/5}\Lambda^3$,
which disagrees with equation \confinings.
For pure $SU(N)$ SYM, the discrepancy worsens.
One computes:
\eqn\puresun{\ev{S(x_1) S(x_2)\ldots S(x_N)}
= {2^N \Lambda^{3N} \over (N-1)! (3N-1)}. }
For large $N$, $\ev S \sim {\Lambda^3\over N}$, in disagreement
with equation \condensaten.

However, some computations do yield
results that agree with the predictions.
In the $SU(2)$ theory with 2 massless flavors,
we find:
\eqn\zerovevsutwo{\ev{{1\over 8} \epsilon^{ijkl}
V_{ij}(x_1) V_{kl}(x_2)} = \Lambda^4}
and for the $SU(N)$ theory with $N$ massless flavors,
we compute using the formalism of
\refs{\germansecond,\cernrev} $\ev{{1\over N!}
\epsilon^{i_1 i_2\ldots i_N}
\epsilon^{\tilde {\imath}_1 \tilde {\imath}_2\ldots
\tilde {\imath}_N} M_{i_1\tilde {\imath}_1}(x_1)
M_{i_2\tilde {\imath}_2}(x_2)
\cdots M_{i_N\tilde {\imath}_N}(x_N)} = {N\over 2N-1}\Lambda^{2N}$
and $\ev{B(x_1) \tilde B(x_2)} = - {N-1\over 2N-1}\Lambda^{2N}$,
which we add to get:
 \eqn\zerovevsun{\ev{\det M-B\tilde B} = \Lambda^{2N}.}
Both equations \zerovevsutwo\ and
\zerovevsun\ are in agreement with the prediction \constraints.
\bigskip \centerline{{\bf Acknowledgements}}

We would like to thank M. Berkooz, K. Intriligator, M. Peskin
and especially N. Seiberg for useful discussions, and
M. Dine, M. Peskin and N. Seiberg for reading a preliminary version
of this paper. We would also like to thank
E. Rabinovici and the organizers
of the Jerusalem Winter School.
This work was supported in part by the Department of Energy under
grant \#DE-AC03-76SF00515 and by a Canadian 1967 Science fellowship.
\listrefs

\end